\documentstyle[prd,aps,floats]{revtex} 
\draft

\begin{document}
\twocolumn[\hsize\textwidth\columnwidth\hsize\csname
@twocolumnfalse\endcsname
\title{
\hbox to\hsize{\large Submitted to Phys.~Rev.~D \hfil E-Print
astro-ph/9610201}
\vskip1.55cm
Primordial Magnetic Fields from Cosmological First Order Phase
Transitions}
\author{G\"unter Sigl and Angela~V.~Olinto}
\address{Department of Astronomy and Astrophysics, Enrico Fermi Instute\\
The University of Chicago,\\ Chicago, Illinois 60637-1433\\
and NASA/Fermilab Astrophysics Center\\
Fermi National Accelerator Laboratory, Batavia, IL 60510-0500}
\author{Karsten Jedamzik}
\address{University of California, Lawrence Livermore National
Laboratory\\
Livermore, CA 94550}
\date{\today}
\maketitle
\begin{abstract}
We give an improved estimate of primordial magnetic fields
generated during cosmological first order phase transitions. 
We examine the charge distribution at the nucleated bubble 
wall and its dynamics. We consider instabilities on the bubble walls 
developing during the phase transition.
It is found that damping of these instabilities 
due to viscosity and heat conductivity caused by particle
diffusion can be important in the QCD phase transition, but is
probably negligible in the electroweak transition. 
We show how such instabilities together with the surface charge
densities on bubble walls excite magnetic fields within a certain range of
wavelengths. We discuss how these
magnetic seed fields may be amplified by MHD effects in the turbulent
fluid. The strength and spectrum of the primordial
magnetic field at the present time for the cases where this mechanism was
operative during the electroweak or the QCD phase transition are
estimated. On a $10\,$Mpc
comoving scale, field strengths of the order $10^{-29}\,$G for
electroweak and
$10^{-20}\,$G for  QCD, could be attained for reasonable
phase transition parameters.

\end{abstract}
\vskip2.2pc]


\narrowtext

\section{Introduction}

The origin of galactic and intra-cluster magnetic fields and the
existence of a cosmological magnetic field are still unknown. 
It has been suggested that dynamo effects
in turbulent fluids might amplify small seed fields
exponentially~\cite{Parker}. However, it has been shown
recently~\cite{VR,KA} that dynamo theory has difficulties 
explaining the galactic and cluster magnetic fields. Alternatively, these
fields may be the result of 
compression of a primordial field which would then permeate the
universe as a whole. In this scenario, the required strength of the
primordial field on a Mpc scale would be much higher, $B_{\rm
prim} \sim 10^{-12}-10^{-9}\,$G at present~\cite{K}, compared to the
case where the fields have been amplified by a dynamo,
$B_{\rm seed} \sim 10^{-24} \,$G. 

It is thus of some interest to investigate mechanisms which could
produce primordial fields and estimate their strength and
spectrum. Scenarios involving inflation can lead to large
coherence lengths, but the predicted amplitude is in general very
small~\cite{TW,R}. Magnetic seed fields can be created by the
slightly different dynamical response of the negatively and
positively charged components of a quasineutral fluid such as in
the Biermann battery mechanism~\cite{Biermann}.
This thermoelectric effect operates as shocks form with the nucleation
of bubbles during the QCD transition~\cite{QLS}, but the created
fields are also quite small. More recently, magnetic seed
fields caused by the turbulent flow near the walls of bubbles which are being
nucleated and grow during a cosmological first order phase
transition have been estimated for the QCD~\cite{CO} and the
electroweak (EW)~\cite{BBL} transition. These two publications make
different assumptions about the nature of the charge layers and the
fluid flow at the bubble walls. As a result,  the seed field
strength estimates in Ref.~\cite{CO} are larger by some 14
orders of magnitude compared to the ones in ref.~\cite{BBL} when 
each generated field is compared to the appropriate equipartition strength for
each transition.

Scenarios where magnetic fields are created directly from the
Higgs fields involved in the EW transition have been discussed
in Refs.~\cite{Vachaspati,EO}. Estimates of magnetic fields
generated during phase equilibration of a complex Higgs field
caused by bubble collisions in a first order phase transition
have been given in Ref.~\cite{KV}.

In this paper, we estimate the strength of primordial
magnetic fields generated during a cosmological first order
phase transition. We rederive the charge density distribution at the
bubble wall in Sec.~II. We then discuss instabilities in
the bubble growth as well as their damping due to a finite shear viscosity
and heat conductivity caused by particle diffusion in Sec.~III.
In Sec.~IV, we show how the fluid flows associated with the 
bubble wall instabilities together with the charge density
distribution at the wall will lead to the generation of magnetic
seed fields. In this section we determine the wavelength range
in which the magnetic fields are initially excited and the amplitude
of these seed fields.
In Sec.~V, we discuss
the transition to the non-linear regime and how
the fluid turbulence can amplify the initial seed fields. We apply
our scenario to the EW and the QCD transition in Sec.~VI and
estimate the resulting strength and spectrum of the extragalactic
magnetic field at present time. A summary of our findings is
contained in Sec.~VII. We will use natural units and the CGS system
throughout the paper.

\section{Surface charge densities at the QCD and the EW phase 
boundaries}

An unmagnetized plasma may develop magnetic fields
if the spontaneous separation of electric charges give rise to a net
current. Separation of charges will occur during a first-order
EW transition through the development of net baryon number or
net top-quark number gradients in or at the EW phase boundaries,
regardless of whether baryogenesis does or does not occur at the
EW transition. During a first order QCD transition net baryon
number discontinuities of appreciable magnitude may develop at the phase
boundaries leading to the spontaneous separation of charges. This
charge separation has been previously considered by Ref.~\cite{CO}
for QCD phase boundaries and by Ref.~\cite{BBL} for EW phase
boundaries.

The first order phase
transition proceeds from a high-temperature (``h'') to a
low-temperature (``l'') phase.
For the EW transition this
would correspond to the phase with a vanishing and non-vanishing
vacuum expectation value of the Higgs field, and for the QCD
transition to the quark-gluon phase and the hadronic phase,
respectively. 

In the QCD transition the baryon number susceptibility in
the high-temperature phase is larger than that in the
low-temperature 
phase such that in chemical equilibrium there is a net baryon 
number density contrast, $R_b=(n_b^h)_{eq}/(n_b^l)_{eq}$, where 
$(n_b^h)_{eq}$ is the net baryon number for the quark-gluon
phase in equilibrium
(high-temperature) and $(n_b^l)_{eq}$ is the same for the hadron phase
(low-temperature).  $R_b \sim 10-10^3$ depending on the transition
temperature. However, chemical equilibrium can only obtain for small
bubble wall velocities, $v_b\ll 1$, and, in general, the net baryon
number contrast can exceed $R_b$ significantly. 

The width of the baryon
number excess layer and the baryon number density contrast between
the quark-gluon phase and the hadron phase can be estimated by studying
the steady-state solutions for the net baryon density in the wall rest
frame. Requiring that the net baryon flux into a thin shell in the
quark-gluon phase at the phase boundary be zero, we find that 
 \begin{equation}
v_N^{th}\Sigma_h n_b^h\biggl({n_b^l\over
n_b^h}-{1\over R_b}\biggr)-{l_q\over L_b}n_b^h+v_b n_b^h=0\,.
\label{bflux}
\end{equation}
The first term in Eq.~(\ref{bflux}) represents the net
flux of baryon
number across  the phase boundary where $v_N^{th}$ is the thermal
nucleon speed in the hadron phase and $\Sigma_h$ is the
probability for a nucleon approaching the phase boundary from
the hadron phase to dissociate
into three quarks and pass over into the quark-gluon phase.
The second and third terms in Eq.(\ref{bflux}) give the diffusive 
baryon flux in
the quark-gluon phase away from the phase boundary and the
advective flux towards the phase boundary, respectively. Here
$l_q$ is the quark mean free path
and $L_b$ is a characteristic length scale of variation in baryon density.
For large bubble velocities, $v_b\sim 0.1$, and/or small baryon number
penetrability, $\Sigma_h\ll 1$~\cite{JF95}, the flux across the phase
boundary is negligible and chemical equilibrium between the phases cannot
obtain. In this case, an exponentially decaying baryon number
layer with width $L_b\simeq l_q/v_b$ develops at the phase
boundary in the quark-gluon phase while far away from the wall baryon number
is homogeneously distributed.  The
out-of-equilibrium density discontinuity at the phase boundary
can then be much larger than its equilibrium value,
$(n_b^h/n_b^l)\gg R_b$. In the opposite limit, e.g. $v_b\ll 1$
and/or $\Sigma_h\simeq 1$, equilibrium is
approximately maintained such that net baryon density is
roughly homogeneous in both phases. Nevertheless, there still
exists a baryon number discontinuity at the
phase boundary with density
contrast $(n_b^h/n_b^l)\simeq R_b$.

For the case of a first order EW transition, let us first
consider the case where baryogenesis occurs at energy scales higher than the
EW breaking scale. In this case a cosmic baryon number asymmetry
(implying a top-quark number asymmetry) already exists in the
high-temperature phase  through,
for example, the existence of an asymmetry in (baryon\, --\, lepton) number
which is conserved by baryon number violating EW sphaleron processes.
Then very similar arguments to net baryon number in the QCD transition hold
for net top-quark number in the EW transition.
Top quark number densities
are thermodynamically suppressed in the low-temperature phase due
to their finite rest mass such that in equilibrium
$(n_t^h/n_t^l)=R_t\simeq 2$. By writing a similar equation to
Eq.(\ref{bflux}) for top quarks at the EW phase boundaries, and
with $v_{top}^{th}\simeq 1$, $\Sigma_{top}\simeq 1$, and
$R_t$ only slightly larger than unity, one finds that only for
relativistic wall velocities, $v_b\simeq 1$, chemical
equilibrium cannot be maintained. Since we consider $v_b\ll
1$ (see below), net top quark number will be in equilibrium which implies
that top quark number density will vary by a factor of order
unity over the extension of the EW bubble wall. 

If EW baryogenesis is
operative during a first order EW transition, the cosmic
baryon asymmetry will be generated either in front of the wall in
non-adiabatic (thin wall) baryogenesis scenarios or within the wall
in adiabatic (thick wall) baryogenesis scenarios~\cite{CKN}. 
Note that for the simplest
extensions of the standard model of EW interactions one
expects thick walls, $L_{\rm wall}\gg 1/T$~\cite{DLHLL}. In this case,
a baryon asymmetry (carried by $u$, $d$, $s$, $c$, and $b$ quarks) 
will smoothly vary
from zero in the high-temperature phase to its cosmic average
value deep within the wall where baryon number violating processes
shut off.

In the QCD transition, the net baryon number is associated with a
net positive charge
density. In particular, $\rho_c=(2/3)en_u-(1/3)en_d$ in the
quark-gluon phase in the limit
where the strange quark mass is large compared to the
temperature, and $\rho_c=en_p$ in the hadron phase where
$n_u$, $n_d$, and $n_p$ are net number densities for
up-quarks, down-quarks, and protons, respectively. Globally
this net positive charge is neutralized by a negative charge
density due to the net electron number. However, locally on the
scale of the baryon discontinuity/inhomogeneity at the phase boundary
electrons are free streaming, e.g. $l_e\simeq[(\alpha^2/T^2)
(g/\pi^2)T^3]^{-1}\gtrsim L_b$, such that the thin excess
baryon number layer/discontinuity at the phase boundary cannot
remain neutral. In the absence of electric fields, $n_u\simeq n_d$
(for $n_e\simeq n_{\nu}\ll n_u$),
while weak equilibrium in the
presence of a positive electric potential requires $n_d>n_u$ and leads to
screening of the charge density by the conversion
of positively charged up-quarks into negatively charged
down-quarks. 

Consider the grand potential
density for up- and down-quarks in the presence of an electric
field, $E$,
\begin{equation}
\Omega=-\sum_{i=u,d}{21\pi^2T^4\over 180}\biggl[1+{30\over
7\pi^2}\biggl({\mu_i- q_i\phi\over T}
\biggr)^2+...\biggr]+{E^2\over 8\pi}\,,\label{EOmega}
\end{equation}
where $\phi$ is the
electrostatic potential, $\mu_i$ and $q_i$ are chemical
potential and electric charge of species $i$, and $\Omega$ is
given to lowest order in $\mu^{\prime}_i= \mu_i-q_i\phi$.
Baryon number density $n_b$ and charge density $\rho_c$ can be
obtained from the
net quark number densities $n_i=-(\partial\Omega
/\partial\mu_i)_{T,\phi}$, which yield
\begin{eqnarray}
n_b&=&{1\over 3}T^2\bigl(\mu_u^{\prime}+\mu_d^{\prime}\bigr)
\,\nonumber\\
\rho_c&=&{1\over3}eT^2\bigl(2\mu_u^{\prime}-
\mu_d^{\prime}\bigr)\,.\label{densities}
\end{eqnarray}
Further demanding weak equilibrium in the case where
$\mu_e=\mu_{\nu}\simeq 0$ (electrons and neutrinos are free
streaming) gives $\mu_d=\mu_u+e\phi$. The Poisson equation together
with Eq.~(\ref{densities}) can then be used to derive the Debye
screening equation
\begin{equation}
{\partial^2\phi\over\partial z^2}-{1\over \lambda^2}\phi
=-2\pi en_b\,,\label{Debye}
\end{equation}
where $\lambda =(4\pi e^2T^2)^{-1/2}$ in this
model. We note here that the actual Debye screening length
in a multi-species relativistic plasma is $\lambda_D\simeq (4\pi
e^2g_cT^2)^{-1/2}$ where $g_c$ is the statistical weight of
relativistic charged particles with mean free path
$l\lesssim\max[\lambda_D,L_b]$.

The Debye equation can be solved for the expected steady state
net baryon number
distribution in the QCD transition and the net top-quark/baryon
number distribution in the 
EW transition. For the
QCD phase boundaries we find a surface charge density
\begin{equation}
\sigma_c\simeq en_b^l\biggl({n_b^h\over
n_b^l}-1\biggr)\lambda_D+{1\over 2}en_b^h
\biggl({\lambda_D\over L_b}\biggr)^2L_b\,.\label{Sigmac1}
\end{equation}
Here the first
term arises from a charge density  on the wall (the width  of
the wall is of order $\sim 1$ fm $\ll \lambda_D$, approximately
a QCD-color screening length) and is spread
over approximately one
electric Debye screening length, whereas the second term results from
a much smaller charge density on the scale of the baryon
number inhomogeneity in the quark-gluon plasma,
$L_b\gg\lambda_D$. The positive charge at the phase boundary in
the quark-gluon plasma is balanced by a negative charge density
at the wall in the hadron phase, resulting in a thin electric
dipole layer. The baryon density at the wall in the quark-gluon
phase, $n_b^h\gg n_b^l$, is approximately $n_b^h\simeq
2\bar{n}_b$ halfway into the transition when chemical
equilibrium is maintained, but can become appreciable towards the
end of the transition, $n_b^h\gg R_b \bar{n}_b$. Here $\bar{n}_b$ is
the cosmic average baryon density, $\bar{n}_b\simeq 40
\eta_S(g/100)T^3$ with $\eta_S\simeq 5\times 10^{-11}$ the
cosmic average baryon-to-entropy ratio.

In the EW transition there are two separate cases
depending on the width of the EW phase boundary, in
particular either thick walls $L_{\rm wall}\gg\lambda_D$, 
or thin walls $L_{\rm
wall}\lesssim\lambda_D$. The resulting surface charge
density, if EW baryogenesis is not operative, is
\begin{eqnarray}
\sigma_c&\simeq&en_{top}^l\biggl({n_{top}^h\over
n_{top}^l}-1\biggr)\biggl({\lambda_D\over L_{\rm wall}}\biggr)^2
L_{\rm wall}\,,\quad\mbox{if $L_{\rm wall}\gg \lambda_D$}
\nonumber\\
\sigma_c&\simeq&en_{top}^l\biggl({n_{top}^h\over
n_{top}^l}-1\biggr)\lambda_D\,,\quad\mbox{if $L_{\rm wall}
\lesssim\lambda_D$}\,,\label{Sigmac2}
\end{eqnarray}
and is seen to be suppressed by
$(\lambda_D/L_{\rm wall})^2$ for continuous thick
walls. Here $n_{\rm top}^h\simeq
(1/6)\bar{n}_b\gtrsim n_{top}^l$ for all but the largest
wall velocities, $v_b\simeq 1$. The magnitude and spatial extension
of charge densities
when EW baryogenesis does occur is of the same order
as that given in Eq.(\ref{Sigmac2}).

To summarize, we parametrize the charge density at the
bubble wall by
\begin{equation}
  \rho_c\sim e\eta T_c^3\,,\label{rhoc}
\end{equation}
which extends over a length scale $l_c\sim f_c/T_c$ around the
wall. In the QCD case $\eta\sim 5\times 10^{-10}-10^{-5}$, 
and $f_c\sim 1$ for the dominant electric charge density 
resulting from the baryon number discontinuity right
at the phase boundary, and
for the EW transition, $\eta\sim 10^{-14}$ and $f_c\sim 40$,
where we assume thick walls, $L_{\rm Wall}\approx 40/T$.

\section{Instabilities of Bubble Growth and Viscous Damping in
the Linear Regime}

During cosmological first order phase
transitions, instabilities may develop as bubbles grow. Hydrodynamic
instabilities can occur when the transport of latent heat is dominated by
the fluid flow. These have been investigated for the QCD
transition~\cite{L} and the EW transition~\cite{KF} in the small
velocities limit, for cosmological detonation
fronts~\cite{AR}, and for general  first order
transitions in the limit of very  small or very large
velocities~\cite{HKLLM}. These bubble wall instabilities may be damped by 
finite viscosity and heat conductivity due to the diffusion of radiation
on small length scales. To account for this damping we will
use  the approach in Refs.~\cite{Weinberg,JKO} for
length scales larger than the radiation mean free path.

We note that, depending on the parameters of the phase
transition, the heat transport may be dominated by diffusion
instead of convection. In that case,  bubbles larger than
the radiation mean free path would likely become unstable to
``dendritic growth''~\cite{FA}. We discuss below under which
conditions this may happen.

The high and low temperature phases for either the EW or the QCD
transition can be described by the following two equations of state
for pressure $p_i(T)$, enthalpy density $w_i(T)$, energy density
$\rho_i(T)$ ($i=h,l$), and temperature $T$:
\begin{eqnarray}
  p_l(T)=[w_l(T)+L]/4\,,&&\quad\rho_l(T)=[3w_l(T)-L]/4
  \,,\nonumber\\
  p_h(T)=w_h(T)/4\,,&&\quad\rho_h(T)=3w_h(T)/4\,,\label{eqstate}
\end{eqnarray}
where $w_i(T)=(2\pi^2/45)g_iT^4$ ($i=h,l$), and $g_i$ is the
number of relativistic degrees of freedom in the
plasma. Furthermore, $\delta\equiv(w_h-w_l)/w_h=L/w_h>0$, with
$w_i\equiv w_i(T_c)$ ($i=h,l$)
and $L$ the latent heat. The critical temperature $T_c$ is
defined by $p_l(T_c)=p_h(T_c)$ or, equivalently, by equating  the
free energy densities in the two phases. We  assume that charged and
strongly interacting particles have approximately the same
velocity ${\bf v}$ and can thus be described in the one-fluid
approximation. The fluid  carries a conserved
quantum number, namely baryon number, with a flux given by
$n^\mu=n U^\mu$. Here, $n$ is the proper baryon number density
and $U^\mu\simeq(1,{\bf v})$ is the four velocity (we assume
non-relativistic flows).
  
In order to obtain the fluid flow and electromagnetic (EM) fields
during bubble nucleation in a first order phase transition, one
would have to solve the following equations in the presence of a
phase boundary simultaneously: Energy-momentum
conservation, $T^{\mu\nu}\,_{;\nu}=0$, where
$T^{\mu\nu}=(\rho+p)U^\mu
U^\nu-pg^{\mu\nu}+\tau^{\mu\nu}+T^{\mu\nu}_{\rm EM}$ is the
energy-momentum tensor including non-ideal contributions
$\tau^{\mu\nu}$ from a
finite viscosity and heat conductivity caused by the diffusion
of photons and neutrinos~\cite{Weinberg} as well as
EM contributions $T^{\mu\nu}_{\rm EM}$; baryon number
conservation, $n^\mu\,_{;\mu}=0$; and, finally, Maxwell's
equations.

A full solution of this set of non-linear equations would require
extensive numerical calculations. Analytical approximations are usually
obtained by linearizing the equations around their zeroth order
equilibrium solution. This has been done in the literature for
various limiting cases: For example, in Ref.~\cite{JKO}, the
case of only one phase was considered and dispersion relations
and damping rates for MHD waves were
obtained. In this work, Maxwell's equations were treated in the
ideal MHD approximation (assuming infinite conductivity which is an
excellent approximation for the evolution of the magnetic field ${\bf
B}$ in the plasma under consideration~\cite{CO}), i.e. the
hydromagnetic equation 
\begin{equation}
\partial_t{\bf B}=\hbox{\boldmath
$\nabla$}\times({\bf v}\times{\bf B})\label{MHD}
\end{equation}
was used.

For an ideal fluid (i.e. vanishing
viscosity and heat conductivity, $\tau^{\mu\nu}=0$) and
no EM fields, the remaining linearized equations
for the perturbations in pressure, $p^\prime$, and velocity,
${\bf v}^\prime$, are~\cite{L,KF}
\begin{eqnarray}
  \left(\partial_t+{\bf v}\cdot\hbox{\boldmath
  $\nabla$}\right)p^\prime+\frac{w}{3}
  \hbox{\boldmath $\nabla$}\cdot{\bf v}^\prime&=&0\label{lineq}\\
  \left(\partial_t+{\bf v}\cdot\hbox{\boldmath
  $\nabla$}\right){\bf v}^\prime+\frac{1}{w}
  \hbox{\boldmath $\nabla$}p^\prime
  -\frac{1}{3}{\bf v}\,\hbox{\boldmath $\nabla$}\cdot{\bf
  v}^\prime&=&0\,,\nonumber
\end{eqnarray}
and have to be solved in both phases. Both the zeroth order
solutions for a planar bubble wall, represented by the
unperturbed quantities ${\bf v}$ and $p$ (or $w$)
and the perturbations ${\bf v}^\prime$ and
$p^\prime$ have to be matched at the phase boundary by requiring
continuity of energy and momentum flow and the transverse
velocity across the interface. The latter condition comes from
assuming equilibrium in the presence of a finite shear
viscosity. It has been shown in
Ref.~\cite{HKLLM} that growing instabilities seem to be possible
only if the bubble wall velocity $v_b$ satisfies
\begin{equation}
  v_b\leq v_{\rm crit}\equiv
  \left[(T_c^2-T_h^2)/2T_c^2\right]^{1/2}\,,\label{instc}
\end{equation}
where $T_h$ is the temperature in the high-temperature phase.
The treatment in Refs.~\cite{L,KF} is only adequate in the
limit $v_b\ll v_{\rm crit}$. It was also
pointed out~\cite{HKLLM} that the QCD transition is a borderline
case since $v_b\simeq v_{\rm crit}\simeq0.03$, whereas the EW transition
is unlikely to fulfill Eq.~(\ref{instc}) in the initial stage of
fast bubble growth. However, since
$v_b\propto(T_c^2-T_h^2)/T_c^2$~\cite{DLHLL}, the condition
Eq.~(\ref{instc}) may eventually be satisfied if $T_h\to
T_c$. Indeed, Ref.~\cite{Heckler} argued  that the
EW bubble walls may slow down significantly (by 1 -- 2 orders of
magnitude) if reheating is significant. For reasonable phase
transition parameters,
$v_b\lesssim10^{-2}$ during this stage, and Eq.~(\ref{instc}) can be met.
It is therefore possible that hydrodynamic instabilities develop both in
the QCD transition and during the late stages of the EW transition for
which the approach in Refs.~\cite{L,KF} can be applied. In the following,
we will assume that this is indeed the case.

Here, we are  interested in solutions of the
general linearized equations for two phases including viscosity
and heat conductivity and EM fields. This is still a quite
complicated problem and  one usually
considers the simpler limiting cases of either one phase of a
viscous fluid or two phases of an ideal fluid. In the general
case, we expect a certain wavelength range over
which there are growing instability modes, i.e. the growing instability
dominates over the diffusive damping. Since we are only
interested in an order of magnitude estimate of the resulting
initial seed fields and their coherence scale, we can restrict
ourselves to the case where the damping rate is small compared
to the instability growth rate. We can then describe these modes by
Eq.~(\ref{lineq}) and verify {\it a posteriori} that the damping
rate $\gamma_{\rm d}$ of the corresponding bubble wall instability is small
compared to its growth rate $\gamma_{\rm inst}$. Seed fields
can then be generated by these growing bubble wall instabilities. We
will verify that backreaction effects of these seed fields onto the
fluid flow in the growing instabilities are negligible also.

Let $v_h$ and $v_l$ be the modulus of the unperturbed velocity
of the high- and low-temperature phase, respectively, in the
bubble wall rest frame. We assume that the first order
transition proceeds as a weak, non-relativistic deflagration,
i.e. $v_h,v_l\ll c_s$, and $\delta\simeq(v_l-v_h)/v_h\ll1$, where
$c_s\simeq1/\sqrt3$ is the speed of sound, and 
get~\cite{L,KF}
\begin{equation}
  \gamma_{\rm inst}\simeq \delta v_hk/2\,.\label{inst}
\end{equation}
This holds true in the range of wavenumbers $k_{\rm min}\lesssim
k\lesssim k_{\rm max}$, with
\begin{eqnarray}
  k_{\rm min}&=&2/R\delta\,,\label{kminmax}\\
  k_{\rm max}&=&\delta w_hv_h^2/\sigma\nonumber\,,
\end{eqnarray}
Here, the first condition comes from requiring that $\gamma_{\rm
d}$  be larger than the expansion rate $v_h/R$ of a
spherical bubble of radius $R$, whereas the second marks the
boundary to the regime where the bubble wall is
stabilized by the surface tension, $\sigma\sim T_c^3$.
The hydrodynamic instability therefore can become
operative for bubbles with a radius $R$ large enough such that
$k_{\rm min}<k_{\rm max}$. This is certainly the case
for the fully developed bubbles near percolation which we 
consider in this paper. For these bubbles, $R\sim R_{\rm
perc}\simeq f_br_H$, where
$r_H\simeq2.3\times10^6(g/10)^{-1/2}(T/100\,{\rm
MeV})^{-2}\,{\rm cm}$ is the Hubble radius, and $g$ is the
number of relativistic degrees of freedom in the
plasma. For the QCD case,
$f_b\sim10^{-6}-10^{-2}$~\cite{FMA,EIKR}, and for the EW
case $f_b\sim10^{-5}-10^{-4}$~\cite{Heckler,EIKR}.

The growth rate in  Eq.~(\ref{inst}) has to be compared with the
damping rate
\begin{equation}
  \gamma_{\rm d}\simeq\frac{g_t}{5g}
  \min\left[k^2\tau_r,\frac{1}{\tau_r}\right]\,,\label{damp}
\end{equation}
where $g_t$ is the statistical weight for particles efficiently
transporting momentum, either neutrinos ($g_t=5.5$) or photons
($g_t=2$), and 
$\tau_r=(\left\langle\sigma_r\right\rangle gT^3/\pi^2)^{-1}$ is
the mean free path of these particles, 
and $\left\langle\sigma_r\right\rangle$ is the
corresponding average cross section. For $k\tau_r\lesssim1$,
Eq.~(\ref{damp}) was obtained from Ref.~\cite{JKO}, and for
$k\tau_r\gtrsim1$, the damping rate is approximately independent
of $k$ and proportional to $1/\tau_r$~\cite{JF}. Associated with
the damping rate of instabilities is the Reynolds number of the
fluid flow at length scale $1/k$,
\begin{equation}
  R(k)\equiv\frac{v_f(k)}{k\tau_r}\,,\label{Reynold}
\end{equation}
where $v_f(k)$ is the typical fluid velocity at that length
scale. In the linear regime
of perturbations, $v_f(k)\lesssim v_h$.

For the EW transition, the dominant processes governing diffusion
of radiation turn out to be EM bremsstrahlung, $e+l\to e+l+\gamma$
and photon-electron pair production,
$e+\gamma\to e+l^+l^-$, where $l$ stands for any charged lepton
in the plasma. These processes are influenced by plasma effects.
With $\alpha$ the fine structure constant, $n_c$ the density of
charged particles in the plasma, and $m_p\simeq(4\pi\alpha
n_c/3T)^{1/2}\simeq0.5T(g/100)^{1/2}$ the plasma mass
of the electron at temperature $T\gg1\,$MeV~\cite{Kapusta}, the
relevant cross sections can be written as
$\left\langle\sigma_r\right\rangle=\left\langle\sigma_{\rm
EM}\right\rangle\simeq(8\alpha^3/m_p^2)
\ln(6.3T/m_p)\simeq3\times 10^{-5}(g/100)^{-1}/T^2$. 
This leads to the Reynolds
number for EM viscosity,
\begin{eqnarray}
  R_{\rm EM}(k)&\simeq&4\times 10^{11}\left({g\over100}\right)^{-1/2}
  \left({T_c\over100{\rm GeV}}\right)^{-1}\nonumber\\
  &&\times f_b\delta v_f(k)\,\frac{k_{\rm min}}{k}\,,\label{REM}
\end{eqnarray}
where $m_{\rm Pl}$ is the Planck mass. Using the numerical
parameters $f_b\gtrsim$ a few $\times10^{-5}$,
$\delta\sim10^{-3}$, and $v_h\simeq v_b=v_l\sim0.01$ in the
percolation stage of the EW
transition~\cite{Heckler,EIKR,IKKL}, we get $R_{\rm EM}(k_{\rm
min})\gtrsim10^2$.

For the QCD transition, the damping rate Eq.~(\ref{damp}) is
dominated by neutrino diffusion for $k\lesssim(\tau_{\rm
EM}\tau_\nu)^{-1/2}$, where $\tau_{\rm EM}$ and $\tau_\nu$ are
the mean free paths of EM radiation and neutrinos,
respectively. Using $\left\langle\sigma_r\right\rangle=
\left\langle\sigma_\nu\right\rangle\simeq2.1G_{\rm F}^2T^2$,
where $G_{\rm F}$ is Fermi's constant, we obtain for the
Reynolds number for neutrino viscosity
\begin{equation}
  R_\nu(k)\simeq10^5\left({g\over10}\right)^{1/2}
  \left({T_c\over100\,{\rm MeV}}\right)^3f_b\delta v_f(k)
  \,\frac{k_{\rm min}}{k}\,.\label{Rnu}
\end{equation}
Using the parameters $f_b\sim10^{-2}$, $\delta\sim10^{-1}$, and
$v_h\sim0.1$~\cite{IKKL} for the QCD transition, we get
$R_\nu(k_{\rm min})\sim1$ and $R_{\rm EM}(k_{\rm
min})\sim10^{12}$.

Combining Eqs.~(\ref{inst}) and (\ref{damp}), we  obtain
\begin{equation}
  \frac{\gamma_{\rm d}}{\gamma_{\rm inst}}\simeq
  \frac{0.2}{\delta v_h}\left({g\over10}\right)^{-1}
  \left({g_t\over 5.5}\right)
  \min\left[k\tau_r,(k\tau_r)^{-1}\right]
  \,,\label{ratio1}
\end{equation}
and instabilities are expected to be damped   in the
wavenumber range
$5(g/10)(g_t/5.5)^{-1}R(k_{\rm min})\lesssim k/k_{\rm min}\lesssim
0.2(g/10)^{-1}(g_t/5.5)R(k_{\rm min})/(\delta v_h)^2$. From this
we see immediately, that, at least in case of the EW transition,
damping is unlikely to be important at the largest length scales
over which instabilities can develop, $k\sim k_{\rm min}$. Damping  
probably only plays a role on length scales which are a few
orders of magnitude smaller than $1/k_{\rm min}$, in a window of
relatively small logarithmic width of $\sim2\log\left[(g_t/5.5)/\delta
v_h(g/10)\right]-1.4$. In contrast, for the QCD transition,
damping could become important around $k_{\rm min}$,
if $R_\nu(k_{\rm min})\lesssim1$.

We note that since $\gamma_{\rm inst}(k_{\rm min})=v_h/R$ is
roughly the convection rate,
$\gamma_{\rm d}(k_{\rm min})/\gamma_{\rm inst}(k_{\rm min})\gtrsim1$
corresponds   to the case where transport of
latent heat is dominated by diffusion instead of hydrodynamic
flow. Therefore, depending on the exact phase transition
parameters entering Eqs.~(\ref{REM}),(\ref{Rnu}), and
(\ref{ratio1}), this situation could occur in the QCD transition, as
remarked in Ref.~\cite{FA}, but not for the EW transition.

\section{Seed Fields from Hydrodynamic Instabilities}

Seed fields associated with  
instabilities will be generated when positive and negative charges are
displaced relative to each other and, therefore, the one fluid approximation
cannot be used. However,   the resulting EM currents,
$j\sim\rho_cv\sim e\eta T_c^3v$, correspond to very small conduction
velocities, $v_c$, of negative relative to positive charge carriers,
$v_c\sim\eta v\ll v$. Therefore, the subsequent evolution of the seed
fields due to dynamo effects can  very well be described within the MHD
approximation. Note that seed fields are necessary since the MHD equation
(\ref{MHD}) is homogeneous in ${\bf B}$. We now estimate
the seed fields generated by  the hydrodynamic
instabilities.

Let the planar,
undisturbed phase boundary be located at $z=0$ with the
low-temperature phase at $z<0$ and the high-temperature phase at $z>0$.
We can choose   
the instability to have a wavevector in the x direction   and the
perturbation of the wall  can be described by
$z_w(x,t)=z_0\exp(\gamma_{\rm inst}t+ikx)$.  This corresponds to a
nonpropagating instability with ${\rm Im}\,\gamma_{\rm inst}=0$. To lowest
non-trivial order in $v_h$ and $\delta$, and for $k\ll k_{\rm
max}$, the solutions of Eq.~(\ref{lineq}) take a relatively
simple form:
\begin{eqnarray}
  {\bf v}^\prime_h&=&\delta v_h\left({\bf e}_z-i{\bf
  e}_x\right)kz_w(x,t)e^{-kz}/2\,,\nonumber\\
  p^\prime_h&=&-\delta w_hv_h^2kz_w(x,t)e^{-kz}/2
  \,,\label{linsol}\\
  {\bf v}^\prime_l&=&\delta v_h\left[{\bf e}_ze^{\delta kz/2}
  +({\bf e}_z+i{\bf e}_x)e^{kz}
  \right]kz_w(x,t)/2\,,\nonumber\\
  p^\prime_l&=&-\delta w_hv_h^2kz_w(x,t)e^{kz}/2
  \,,\nonumber
\end{eqnarray}
where ${\bf e}_x$ and ${\bf e}_z$ are unit vectors in the $x$ and
$z$ direction, respectively. Note that Eq.~(\ref{linsol}) only
holds for distances from the wall which are much larger than the
perturbation amplitude of the wall location, i.e. $z\gg z_w(x,t)$. It is
seen from Eq.~(\ref{linsol}) that the instability perturbs the fluid flow
in the low-temperature phase up to a characteristic distance $1/\delta k$,
whereas in the high-temperature phase the fluid flow is only perturbed up to
a distance $1/k$. For the wavelength regime under consideration,
Eq.~(\ref{kminmax}), these scales are typically much larger than the width
$\sim f_c/T_c$ of the dipole layer.

Eq.~(\ref{linsol}) illustrates that, for a given mode with
${\bf k}=k{\bf e}_x$, the $x$ component of the
perturbed velocity field has a discontinuity at $z=0$:
\begin{equation}
  \left({\bf v}^\prime_h-{\bf v}^\prime_l\right)\cdot{\bf e}_x=
  -i\delta v_hkz_w(x,t)\,.\label{disc}
\end{equation}
The dipole layer with charge density given by Eq.~(\ref{rhoc})
thus leads to a net current in the $x$ direction,
\begin{equation}
  j_x\sim e\eta T_c^3\delta v_hkz_w(x,t)e^{-T_c\vert z\vert/f_c}
  \,,\label{jx}
\end{equation}
provided $f_c/T_c\gtrsim z_w(x,t)$. We can
now superpose all instability modes from Eq.~(\ref{linsol}) with
${\bf k}$ in the $z=0$ plane and write the wall displacement
spectrum as $\left\langle z_w^2\right\rangle=\int[z_w(k)]^2d\ln
k$. Using Amp\'{e}re's law without the displacement current,
$\hbox{\boldmath $\nabla$}\times{\bf B}=4\pi{\bf j}$, and
Eq.~(\ref{jx}), after performing a three dimensional Fourier
transformation, we arrive at
\begin{equation}
  B_s(k)\sim 4e\eta T_c^2\delta v_hf_c
  \left(k_{\rm min}k\right)^{1/2}\vert z_w(k)\vert\label{Bs1}
\end{equation}
for the seed field spectrum $\left\langle
B_s^2\right\rangle=\int[B_s(k)]^2d\ln k$. Strictly speaking,
Eq.~(\ref{Bs1}) is only reliable for $z_w(k)\lesssim f_c/T_c$.

The instabilities Eq.~(\ref{linsol}) grow non-linear when the
perturbed velocity spectrum
\begin{equation}
  v_f(k)\sim \delta^{1/2}v_h\left(k_{\rm min}k\right)^{1/2}
  \vert z_w(k)\vert\label{vfk}
\end{equation}
satisfies $v_f(k)\gtrsim v_h$, where, analogous to the
seed field spectrum, the flow velocity spectrum is written as
$\left\langle v_f^2\right\rangle=\int[v_f(k)]^2d\ln k$.
Furthermore, the seed field ${\bf
B}_s$ and its source currents ${\bf j}$ give rise to
an additional force acting onto the
fluid of the form ${\bf j}\times{\bf B}$. This
term is of second order in the perturbation amplitude and was
neglected in Eq.~(\ref{lineq}). We now compare its magnitude with the
leading force term $\hbox{\boldmath $\nabla$}p^\prime$ which is
of first order in the perturbation. By using
Eqs.~(\ref{linsol}), Amp\'{e}re's law and an equation analogous
to Eq.~(\ref{vfk}) for $\vert\hbox{\boldmath
$\nabla$}p^\prime\vert$, we obtain
\begin{equation}
  \frac{\vert jB_s\vert}
  {\vert\hbox{\boldmath $\nabla$}p^\prime\vert}
  \simeq\frac{(e\eta f_c)^2}{v_h}\frac{T_c^4}{w_h}
  \delta^{1/2}v_f(k)\,,\label{force1}
\end{equation}
on a length scale $1/k$. This is small compared to unity in the linear
regime where $v_f(k)\lesssim v_h$. In addition, the
electric field $E_s(k)$ induced by the growing magnetic seed
field, $E_s(k)\sim[\partial B_s(k)/\partial t]/k\sim\delta v_h
B_s(k)$, is small compared to the zeroth order field $E\sim
e\eta f_cT_c^2$ caused by the electric charge layer,
$E_s(k)/E\sim v_f(k)\ll1$.
We therefore conclude that in most cases backreaction
effects of the seed fields onto the medium are negligible in the
linear regime. Therefore, a  conservative estimate of the seed
field $B_s$  is  given by assuming $z_w(k)\sim f_c/T_c$;
\begin{equation}
  B_s(k)\sim 4e\eta f_cT_c^2\delta^{1/2}v_f(k)\gtrsim
  4e\eta T_c^2\delta v_hf_c^2
  \left(\frac{k_{\rm min}k}{T_c^2}\right)^{1/2}
  \,.\label{Bs2}
\end{equation}
Being less conservative, one can extrapolate the first
expression in Eq.~(\ref{Bs2}) up to the
transition to the non-linear regime, where $v_f(k)\simeq v_h$,
and obtain the optimistic estimate $B_s(k)\sim e\eta
T_c^2\delta^{1/2}v_hf_c$. In the next section we will show that once the
non-linear regime is reached, enough turbulence in the fluid is expected
to build up and the seed fields can be amplified by dynamo effects.

For comparison, the thermoelectric effect discussed in
Ref.~\cite{QLS} for the QCD transition leads to seed fields of
the order $\sim0.01v_fT_c/R_{\rm perc}$ which is smaller than
Eq.~(\ref{Bs2}) for $k\gg k_{\rm min}$. This can be seen by
rewriting Eq.~(\ref{Bs2}) as $B_s(k)\gtrsim4\eta
v_hf_c^2(T_c/R_{\rm perc})(k/k_{\rm min})^{1/2}$.
Other mechanisms have been investigated in the literature starting
from thermal field fluctuations on a length scale $\simeq1/T_c$,
$B_{th}^2(k\simeq T_c)\simeq8\pi T_c^4$. On larger scales,
$k<T_c$, the magnetic field spectrum was argued to behave as a
power-law, $B_{th}(k)\simeq(8\pi)^{1/2}T_c^2(k/T_c)^p$. Several
options have been discussed for $p$: Based on stochastic
arguments applied to the order parameter involved in the phase
transition, Ref.~\cite{Vachaspati} argued for $p=1$, and
Ref.~\cite{EO} for $p=1/2$. In contrast, treating the
resulting magnetic dipole moments as stochastic variables leads
to $p=3/2$~\cite{Hogan}. In any case, by comparing with
Eq.~(\ref{Bs2}) we see that for $p\gtrsim1$, very roughly, on a
scale $k\simeq k_{\rm min}$ given by Eq.~(\ref{kminmax}), our
mechanism leads to comparable or stronger fields already in the
linear regime.

\section{Amplification of Seed Fields by MHD Effects}

The seed fields discussed in the previous section can be
amplified by exchange of energy with the turbulent fluid
flow. This non-linear MHD regime  is usually investigated by
numerical simulations~\cite{dyn}. The magnetic field energy
$E_M$ typically grows up to equipartition with the fluid flow
where it takes the value
\begin{equation}
  E^{eq}_M=\frac{B_{eq}^2}{8\pi}=gT^4v_f^2\,,\label{equi}
\end{equation}
with $B_{eq}$ being the corresponding equipartition field.
The fluid velocity in the turbulent regime can be estimated by
assuming that the latent heat release $L\simeq\delta w_h$ is at
least partially converted into turbulent motion.
This results in $v_f\simeq\delta^{1/2}$ which is probably
somewhat larger than
$v_f\lesssim v_h$ in the linear regime. 

We note in passing that the  
resulting strong equipartition fields $B_{eq}$ on bubble scales
do not cause a problem with recent bounds from big-bang
nucleosynthesis~\cite{BBN}. This is because fields on scales
smaller than the comoving horizon size at neutrino-decoupling
are expected to be significantly damped by neutrino
diffusion by the time nucleosynthesis commences~\cite{JKO}.

The required amount of amplification for the seed field in
Eq.~(\ref{Bs2}) is given by the inverse of the ratio
\begin{eqnarray}
  \frac{B_s(k)}{B_{eq}}&\simeq&0.1\eta
  f_cv_f(k)\left({g\over10}\right)^{-1/2}
  \nonumber\\
  &\gtrsim&10^{-17}\left({T_c\over100\,{\rm GeV}}\right)
  \frac{\eta f_c^2}{f_b\delta^{1/2}}\,v_h\,,\label{amp}
\end{eqnarray}
which is $\gtrsim 10^{-24}(\eta/10^{-14})$ for the EW and
$\gtrsim 10^{-24}(\eta/10^{-5})$ for the QCD transition, for the
parameters used above.
Note, that in case of the EW transition,
$B_s/B_{eq}$ is much larger than what was estimated in
Ref.~\cite{BBL} if it is scaled to the same $\eta$. The reason
is that our instability analysis
revealed the existence of a monopole layer of an EM current on
the bubble wall, whereas in Ref.~\cite{BBL} a dipole layer was
assumed.

The central question in the theory of magnetic field
amplification in a turbulent fluid concerns the time scale over
which energy equipartition is achieved and the spectrum of the
resulting magnetic field. Here we use the analytic approach
and  physical arguments 
developed in Ref.~\cite{KA} which was recently applied to
the problem of galactic field generation~\cite{KCOR}. Once the
bubbles start to collide, vorticity is expected to develop
which, on a scale $k$ is of the order $kv^\prime(k)$ and
corresponds to a turnover rate of equal size. This
turnover rate becomes larger than the instability growth rate
Eq.~(\ref{inst}) once $v^\prime(k)\gtrsim\delta v_h$. If the
Reynolds number $R(k)\gg1$, one
therefore expects the fluid flow to turn turbulent before the
instabilities become non-linear, which occurs for
$v^\prime(k)\simeq v_h$. At that point, as long as the
velocities involved are smalled compared to the speed of sound
$c_s\simeq1/\sqrt3$, one expects a
Kolmogorov type of spectrum to develop. The typical
velocity on a scale $k$ is then given by
$v_f(k)\simeq v_f(k_c/k)^{1/3}$~\cite{Landau}. This holds for
$k_c\lesssim k\lesssim k_{\rm vis}\simeq k_cR(k_c)^{3/4}$,
where $k_c$ is the wavenumber for which the eddy turnover rate
$\simeq k_cv_f$ is equal to the inverse of the percolation
time $t_{\rm perc}\simeq R_{\rm perc}/v_h$, giving
$k_c\simeq v_h/R_{\rm perc}v_f\simeq1/R_{\rm perc}$. The viscous
scale $k_{\rm vis}$,
at which dissipation becomes important, is basically the scale
of the smallest eddy in the turbulent flow.

Neglecting resistive damping for the moment, the growth rate of
the magnetic field energy during the MHD regime is
given by~\cite{KA}
\begin{equation}
  \frac{dE_M}{dt}=2\gamma_ME_M\,,\label{dEM}
\end{equation}
with
\begin{equation}
  2\gamma_M\simeq\int v_f(k)dk\,,\label{gammaM}
\end{equation}
where the velocity spectrum is again written as $\left\langle
v_f^2\right\rangle=\int[v_f(k)]^2d\ln k$. Thus, $\gamma_M$ is
basically the maximum turnover rate
of the eddies in the flow. For a Kolmogorov spectrum,
\begin{equation}
  \gamma_M\simeq v_fk_c\left({k_{\rm vis}\over k_c}\right)^{2/3}
  \simeq v_fk_c\left[R(k_c)\right]^{1/2}\,.\label{gammaMK}
\end{equation}
The resistive scale $k_r$ is defined as the wavenumber where the
resistive damping rate $\simeq-k^2/(4\pi\sigma_{\rm cond})$
caused by the finite conductivity $\sigma_{\rm
cond}\simeq10T_c$~\cite{Ahonen} becomes comparable to the growth
rate $2\gamma_M$. Thus, $k_r\simeq(4\pi\sigma_{\rm
cond}\gamma_M)^{1/2}\simeq k_{\rm vis}[R_M/R(k_c)]^{1/2}$, where
$R_M=4\pi(2\pi/k_c)v_f\sigma_{\rm
cond}\simeq10^5\left(g/100\right)^{-1}R_{\rm EM}(k_c)$ is the
magnetic Reynolds number. The magnetic field spectrum cuts off
exponentially for $k\gtrsim k_r$.
It has been shown in Ref.~\cite{KCOR} that $\gamma_M$ is
constant and $E_M$ grows exponentially until $2\gamma_ME_M$
becomes comparable with the turbulent power $P\simeq gT_c^4\int
v_f^3(k)dk\simeq gT_c^4k_cv_f^3\ln R(k_c)$ which is also the
dissipation rate due to viscosity. At that point, the Alfv\'en
velocity $v_A\simeq(2E_M/gT_c^4)^{1/2}$ becomes comparable to
the turbulent velocity at the viscous scale, $v_f(k_{\rm
vis})\simeq v_fR(k_c)^{-1/4}$, which implies $E_M\simeq
E_M^{eq}R(k_c)^{-1/2}$. Since the magnetic field on scales $k$
with $kv_A\gtrsim\gamma_M\simeq k_{\rm vis}v_f(k_{\rm vis})$
cannot grow, the magnetic field energy should be concentrated
around $k_{\rm vis}$ by the time when
$P\simeq2\gamma_ME_M$. From then on, $2\gamma_ME_M\simeq P$
is roughly constant, so that $E_M$ only grows linearly in time
until it reaches $E_M^{eq}$. Therefore, at the same time,
$\gamma_M$ has to decrease until it reaches $\simeq v_fk_c$. By
comparing with the second equality in Eq.~(\ref{gammaMK}), this
means that the high
wavenumber cutoff in the velocity power spectrum decreases until
it reaches $k_c$. The viscous length scale grows due to the extra
viscosity caused by the energy drain from the smallest eddy to
the magnetic field. The same happens to the magnetic field
spectrum, so that in the end all the magnetic energy $E_M^{eq}$
is concentrated at $k=k_c$, i.e. at the largest wavelengths. The
time scale over which this process occurs can be estimated as
\begin{equation}
  t_M\simeq\int{dE_M\over2\gamma_ME_M}\simeq{E_M^{eq}\over P}
  \simeq\frac{1}{k_cv_f}\simeq t_{\rm perc}\,.\label{T}
\end{equation}
We have thus verified that amplification up to equipartition is
possible within the percolation time. Furthermore, the magnetic
field is expected to be coherent on the scale $k_c$, and we can
write $B(1/k_c,T_c)\simeq(8\pi g\delta)^{1/2}T_c^2$.

\section{Relic Extragalactic Magnetic Fields}

On length scales $r\gtrsim1/k_c$ we will again assume a
power-law behavior $\propto(k_cr)^{-p}$ for the magnetic
field. Since $k_c$ is a macroscopic scale
compared to the correlation length of the order parameter of the
phase transition which is $\sim1/T_c$, the large scale fields
should be determined by the randomly oriented magnetic dipole
moments, leading to $p\simeq3/2$~\cite{Hogan,CO,BBL}. For the
magnetic field on a comoving scale $r$ at redshift $z$ this
eventually leads to
\begin{eqnarray}
  B(r,z)&\sim&10^{-21}(1+z)^2f_b^{3/2}\delta^{1/2}
  \left({g\over100}\right)^{-1/4}\nonumber\\
  &&\times\left({T_c\over100\,{\rm GeV}}\right)^{-3/2}
  \left({r\over10\,{\rm Mpc}}\right)^{-3/2}
  \,{\rm G}\,.\label{Bls}
\end{eqnarray}

We could have followed previous authors by choosing a fiducial value
for $r$ of 1 Mpc, which corresponds to the scale that collapsed to form a
galaxy. The estimated  field would then  be appropriate for 
seeding the galactic fields. However, it was recently suggested that
magnetic fields on scales smaller than the Silk scale $\simeq10\,$Mpc are
likely to be damped  by photon diffusion at recombination~\cite{JKO}.
Thus, Eq.~(\ref{Bls}) probably only applies for $r\gtrsim10\,$Mpc and
leads to an extragalactic magnetic
field spectrum $B(r,z=0)\sim10^{-29}(r/10\,{\rm
Mpc})^{-3/2}\,$G for the EW transition, and
$B(r,z=0)\sim10^{-20}(r/10\,{\rm Mpc})^{-3/2}\,$G for the
QCD transition for the parameters used above. In the latter
case our mechanism thus predicts a field strength on a $10\,$Mpc
scale which is roughly
comparable to the field expected from the Biermann battery
mechanism acting on intergalactic distance scales during
large-scale structure formation~\cite{KCOR}.

For comparison, the stochastic models mentioned in Sec.~III
lead to a field strength
$B(r,z)\simeq(8\pi)^{1/2}T_0^2(1+z)^2(rT_0)^{-p}$, where
$T_0\simeq2.7\,$K is the present temperature of the cosmic
microwave background. For both the EW and the QCD transition
this gives $B(10\,{\rm Mpc},z=0)\sim10^{-19}\,$G and
$\sim10^{-32}\,$G for $p=1/2$~\cite{EO} and
$p=1$~\cite{Vachaspati}, respectively.

In Refs.~\cite{BEO,DD} it was pointed out that hydromagnetic
turbulence in the early universe might shift a given magnetic
field power spectrum near equipartition with the fluid to
length scales larger by a factor $f$ which can be as large as a
few orders of
magnitude. In that case, the length scale $r$ in the above
estimates should be substituted by $r/f$. At large scales
characterized by a given $r$, this could increase the fields by
a few orders of magnitude.


\section{Conclusions}

We have discussed the generation of magnetic fields during the
growth of weak deflagration bubbles nucleated in a cosmological
first order phase
transition in some detail. This proceeds essentially in two
steps: First, instabilities can arise which can be described by
linear perturbation theory. Together with a finite charge
density along the bubble walls these instabilities cause EM
currents and thus magnetic seed fields. While  currently we are unable
to decide unambiguously whether such instabilities  
form,  for reasonable phase transition
parameters this is at least plausible. Taking into account damping
due to a finite viscosity and heat conductivity
caused by radiation diffusion we showed that the resulting bubble
instabilities can grow on length scales somewhat smaller than the
bubble radius at percolation. Once these perturbations
grow non-linear, the fluid is expected to turn turbulent and the
seed fields can be amplified by MHD effects in the form of a
dynamo. Qualitative physical arguments show that equipartition
of the magnetic field energy with the kinetic energy in the
turbulent motion can be achieved and the field spectrum can be
concentrated at length scales not much smaller than the bubble
radius at percolation. Random superposition of the magnetic
dipole moments associated with the bubbles leads to large scale
fields which can be estimated and have a characteristic
spectrum. On a $10\,$Mpc comoving scale, field strengths of the order
$10^{-29}\,$G from the EW transition and $10^{-20}\,$G from the
QCD transition could be attained at $z=0$ for reasonable phase
transition parameters. Finally, hydromagnetic turbulence after the phase
transition may further enhance the large scale fields.


\section*{Acknowledgments}

We are indebted to Gordon Baym, Kari Enqvist, Michal Hanasz,
Russell Kulsrud, Harald Lesch, Robert Rosner, and Tanmay
Vachaspati for many
interesting and helpful discussions on the subject of
cosmological magnetic fields.
This work was supported by the DoE, NSF, and NASA at the
University of Chicago, 
by the DoE and by NASA through grant NAG 5-2788 at Fermilab, and
by the Alexander-von-Humboldt Foundation. We thank the Aspen Center for
Physics for the hospitality while some of this work was completed.


\end{document}